\documentclass[conference]{IEEEtran}

\usepackage[utf8]{inputenc}
\usepackage[T1]{fontenc}
\usepackage{cite}
\usepackage{amsmath,amssymb}
\usepackage{array}
\usepackage{graphicx}
\usepackage{url}
\usepackage{booktabs}
\usepackage{balance}
\usepackage{xcolor}
\usepackage{listings}
\usepackage{balance}
\usepackage{hyperref}
\title{Gateway-Free Kafka Streaming from FPGA-Based Detectors:\\
A Bare-Metal Pipelined Producer on Zynq-7000}

\author{
\IEEEauthorblockN{
Fernanda Zapata Bascu\~n\'an
}
\IEEEauthorblockA{
Universidad Nacional de la Defensa (UNDEF)\\
C\'ordoba, Argentina\\
\texttt{fzapata541@alumnos.iua.edu.ar}
}
\and
\IEEEauthorblockN{
Mat\'ias Rolf Hampel
}
\IEEEauthorblockA{
Universidad Tecnol\'ogica Nacional (UTN)\\
Buenos Aires, Argentina\\
\texttt{matias.hampel@iteda.gob.ar}
}
}

\begin{document}
\maketitle

\begin{abstract}
We present a bare-metal Apache Kafka producer implemented on a
Zynq-7000 System-on-Chip (SoC) for event-driven data acquisition. Events are
generated in Programmable Logic (PL), transferred to the Processing System
(PS) through AXI4-Stream and AXI DMA, and published to a standard Kafka
broker without an intermediate gateway computer. The PS constructs Kafka
\texttt{ProduceRequest} messages and transmits them through the lwIP raw TCP/IP
API. A configurable pipeline limits the number of outstanding Kafka requests.

Pipeline depths $D\in\{1,2,4,8\}$ were evaluated with two independent DMA
sources. For each depth, 20 independent runs of 5000 events were acquired,
for 400\,000 event records in total. Each event carries a 32-bit frame counter
and a 32-bit PL timestamp. PS-side instrumentation records DMA start and
completion, publication, successful local \texttt{tcp\_write()} acceptance,
and reception of the corresponding Kafka acknowledgement. Measurement
records are exported to a separate Kafka topic after the timed event stream
has completed.

Each of the 80 runs contained 5000 records with frame values from 0 to 4999,
with no sequence discontinuities in the exported metric records and no
recorded Kafka errors. Mean operational throughput was approximately
5.3~kevents/s at $D=1$, 9.4--9.6~kevents/s at $D=2$, 16.2~kevents/s at
$D=4$, and 16.8--17.5~kevents/s at $D=8$. The median
publish-to-acknowledgement interval was 1.62--1.64~ms at $D=1$,
0.98~ms at $D=2$, 0.66~ms at $D=4$, and 0.83--0.86~ms at $D=8$.
At $D=4$, the corresponding P95 values were 0.81--0.83~ms.
\end{abstract}

\begin{IEEEkeywords}
Apache Kafka, bare-metal systems, Zynq-7000, FPGA, lwIP, data acquisition,
event-driven readout, embedded networking.
\end{IEEEkeywords}

\section{Introduction}

Apache Kafka~\cite{kafka_original} stores records in ordered topic partitions
and allows consumers to read or replay them from a selected offset. These
properties provide persistent storage and per-partition ordering for binary
data-acquisition (DAQ) records.

Standard Kafka clients normally run on systems with an operating system,
socket APIs, threads, and dynamic memory management. These assumptions do not
map directly onto a small bare-metal FPGA/SoC application. The implementation
presented here places the Kafka producer on the same Zynq-7000 SoC that
receives the FPGA-generated events, so the measured path does not include a
separate gateway computer.

The producer implements the Kafka Wire Protocol on the Zynq PS using the lwIP
raw API, statically allocated buffers, and an explicit finite-state machine
(FSM). Events generated in the PL are transferred through AXI4-Stream and AXI
DMA, serialized into Kafka Record Batches, and sent to a remote broker over
Gigabit Ethernet.

Implementing this path without an operating system requires the application to
construct the Kafka binary request format directly, including Record Batch
serialization, variable-length fields, CRC-32C generation, request correlation,
and response parsing. TCP progress is driven by the lwIP raw callback API, so
session state and request ownership remain explicit in the application rather
than being delegated to a client library or worker thread.

The measurement method uses producer-side timing. Each event contains a PL
frame counter and hardware timestamp. The PS records timestamps at DMA start,
DMA completion, publication, successful local \texttt{tcp\_write()}
acceptance, and reception of the associated Kafka \texttt{ProduceResponse}.
These timestamps define the DMA, software-queue, send-to-acknowledgement, and
publish-to-acknowledgement intervals used in the analysis. The PL and PS time
values are not subtracted from each other because a common clock origin has
not been established.

The implementation is evaluated for pipeline depths $D\in\{1,2,4,8\}$ with
two independent DMA sources. The dataset contains 80 independent runs and
400\,000 acknowledged event metric records. Throughput and latency statistics
are calculated at run level, with confidence intervals across independent
runs.

\section{Background and Related Work}
\label{sec:background}

\subsection{Kafka Wire Protocol}

Apache Kafka uses a binary TCP-based protocol~\cite{kafka_protocol}. Each
request begins with a length-prefixed request frame whose header identifies the
API, API version, correlation identifier, and client. The producer in this work
uses \texttt{ProduceRequest} version~3. The request body identifies the target
topic and partition and contains a Record Batch carrying the application
payload.

Record Batch format version~2 combines fixed-width fields with variable-length
record fields and a CRC-32C checksum over the batch fields following the magic
byte~\cite{crc32c_castagnoli}. The implementation therefore performs Kafka
serialization separately from the 8-byte DAQ payload: the FPGA event remains
unchanged while the PS constructs the surrounding protocol frame.

The correlation identifier is also used for request concurrency. Several
\texttt{ProduceRequest}s may remain outstanding on one TCP connection, and the
producer uses the identifier in each \texttt{ProduceResponse} to release the
matching in-flight slot rather than assuming responses correspond to the most
recently submitted event.

\subsection{lwIP on Zynq Bare Metal}

The lwIP stack~\cite{lwip_dunkels} is available in the AMD-Xilinx embedded
software environment~\cite{xilinx_lwip}. Its raw API is callback based and
does not require a POSIX socket layer. The application services Ethernet
input explicitly, and TCP callbacks execute when the stack processes the
corresponding network events.

The producer therefore keeps connection establishment, Kafka session setup,
request transmission, response parsing, and error handling in explicit
software state. The main loop continues to service the network while a Kafka
request is outstanding.

\subsection{DAQ Context}

FPGA-based front ends are used for local event formation, timestamping,
buffering, and readout in detector systems. Examples include distributed
instrumentation in astroparticle physics~\cite{auger_amiga} and proposed
underground instrumentation for ANDES~\cite{andes_detector}. The present work
addresses the narrower implementation problem of publishing FPGA-acquired
records directly from a bare-metal Zynq PS to Kafka. No comparison with a
Linux Kafka client is made in the experimental results.

For clarity, Table~\ref{tab:path_comparison} distinguishes the measured
architecture from a gateway-based Kafka path. The table describes the location
of functions in the two architectures; it is not a performance comparison.

\begin{table}[!t]
\renewcommand{\arraystretch}{1.08}
\caption{Kafka Publication Path at the FPGA Front End}
\label{tab:path_comparison}
\centering
\scriptsize
\setlength{\tabcolsep}{2pt}
\begin{tabular}{@{}p{0.33\columnwidth}p{0.24\columnwidth}p{0.28\columnwidth}@{}}
\toprule
Property & Gateway-based path & Present implementation\\
\midrule
Kafka producer & External OS host & Zynq PS\\
FPGA-to-producer transfer & External/local interface & AXI DMA\\
OS in producer path & Yes & No\\
Dynamic allocation in measured producer & Client dependent & No\\
Kafka Wire Protocol on acquisition SoC & No & \textbf{Yes (This work)}\\
\bottomrule
\end{tabular}
\end{table}

\section{System Architecture}
\label{sec:architecture}

\begin{figure}[!t]
\centering
\includegraphics[width=\columnwidth]{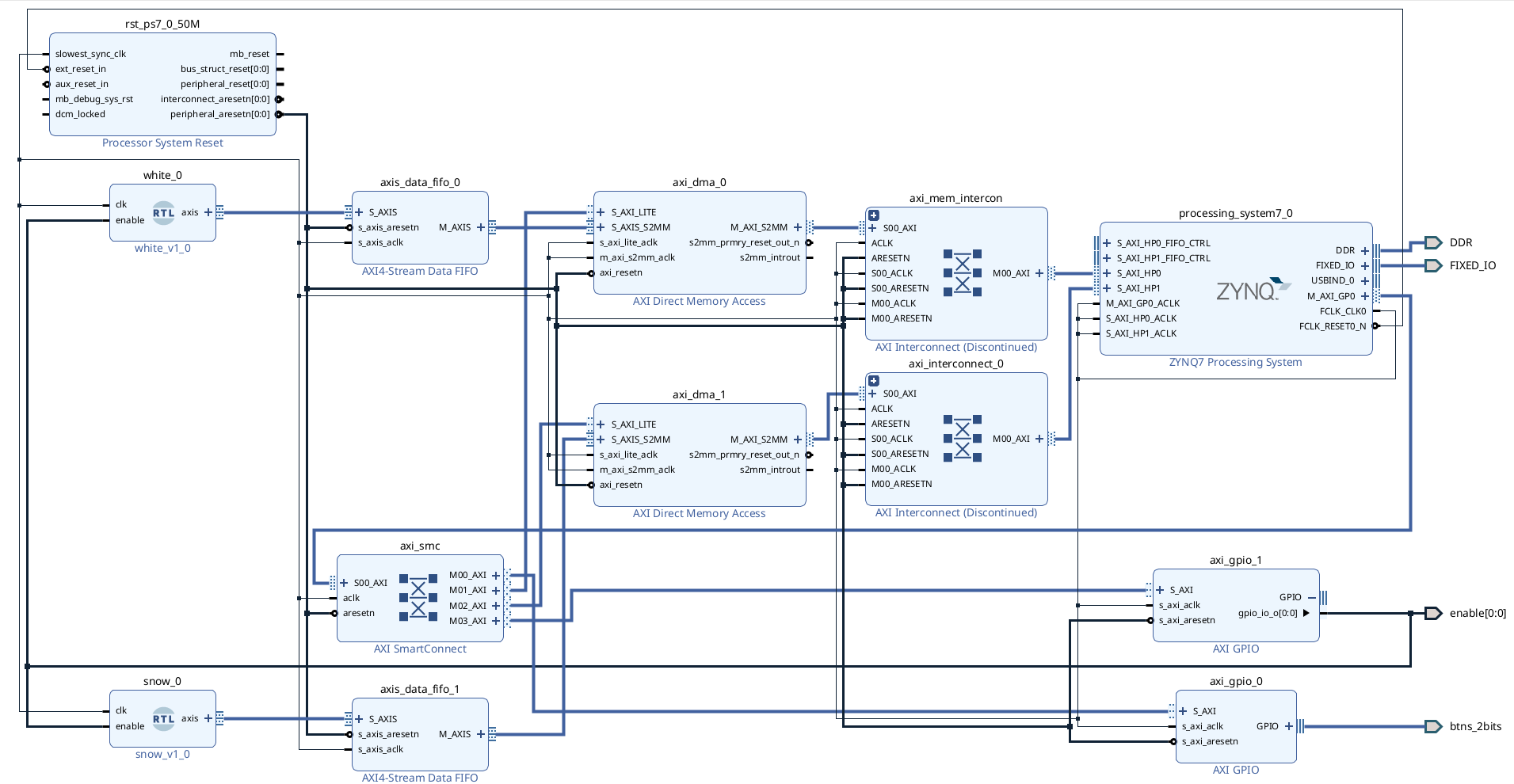}
\caption{System architecture. Two PL event sources use independent AXI DMA
paths to the bare-metal PS application. The PS constructs Kafka
\texttt{ProduceRequest} messages and communicates with the broker over Gigabit
Ethernet.}
\label{fig:architecture}
\end{figure}

The experimental platform is a Digilent Cora Z7-07S containing a Zynq-7000
SoC~\cite{cora_refman}. The PL contains the event sources and AXI4-Stream data
paths. The PS runs the DMA acquisition code, lwIP, and the Kafka producer. The
Ethernet link operates at 1~Gb/s.

\subsection{Programmable Logic}

Two independent event sources are implemented in the PL and connected to the
PS through separate AXI DMA paths. Each event consists of one 64-bit
AXI4-Stream transfer. The upper 32 bits contain a monotonically increasing
frame counter and the lower 32 bits contain the PL timestamp:

\begin{equation}
\texttt{TDATA}[63:0] =
\{\texttt{frame}[31:0],\,\texttt{timestamp}[31:0]\}.
\label{eq:event_format}
\end{equation}

For each event, \texttt{TKEEP=0xFF} and \texttt{TLAST=1}. The Kafka
application payload is therefore 8~bytes. The measurement records identify
the two acquisition paths as source~0 and source~1.

The PL timestamp is retained in raw counter units. The PL timestamp and the
PS timing functions do not share a demonstrated common epoch, so the PL value
is not subtracted from PS timestamps in the latency calculations.

\subsection{Processing System}

The PS runs the Xilinx standalone bare-metal environment. The application
uses the lwIP raw API and a cooperative main loop. During acquisition, the
loop services Ethernet input, protocol timers, Kafka state transitions, and
the selected DMA channel. No RTOS or blocking socket interface is used.

Both AXI DMA instances operate in simple S2MM mode. Each transaction receives
one 8-byte AXI4-Stream event into a 32-byte-aligned buffer. Cache maintenance
is performed around the receive operation: the buffer is prepared before the
S2MM transfer and invalidated after completion before the CPU reads the received
word. The frame counter and PL timestamp are then extracted from the 64-bit
value.

The main loop does not busy-wait when the Kafka producer is full. Before arming
the next DMA transfer, it checks whether an in-flight slot or software-queue
entry is available. Ethernet input and protocol timers continue to be serviced
while producer capacity is unavailable. This keeps network progress active
while back-pressure is applied to event acquisition.

For each acknowledged event, the PS stores
\texttt{dma\_start\_us}, \texttt{dma\_done\_us}, \texttt{publish\_us},
\texttt{send\_us}, and \texttt{ack\_us}. The value \texttt{send\_us} is
recorded after \texttt{tcp\_write()} accepts the request data. It marks local
lwIP acceptance, not physical transmission on Ethernet. The value
\texttt{ack\_us} is recorded when the receive callback processes the
corresponding Kafka \texttt{ProduceResponse}.

\section{Kafka Wire Protocol Implementation}
\label{sec:protocol}

\subsection{Static Memory Model}

The publication path does not use dynamic allocation. Request buffers,
response storage, in-flight request descriptors, and the software queue are
allocated statically, with sizes fixed at compile time. Each outstanding
request retains its transmit storage until the matching Kafka response is
processed. This prevents a buffer from being reused while a request remains
in flight.

\subsection{Serialization and CRC-32C}

Kafka request-level fields use network byte order, while Record Batch records
also contain variable-length encoded fields. The implementation provides local
routines for fixed-width big-endian values and signed varints. The 8-byte DAQ
payload is inserted as the record value; Kafka framing is constructed around
that value in PS software.

CRC-32C is computed in software with a lookup table initialized at startup.
The checksum covers the Record Batch fields specified by the Kafka
protocol~\cite{kafka_protocol,crc32c_castagnoli}. Request construction is
therefore performed without an external serialization library.

\subsection{Asynchronous Session Management}

The Kafka session is represented as an FSM driven by lwIP callbacks. After TCP
connection establishment and Kafka session setup, the producer enters the
ready state. Incoming TCP data are parsed as Kafka responses, and the
correlation identifier is used to locate the corresponding in-flight request
slot. Processing that response records the acknowledgement timestamp and
releases the slot for a later event.

Because lwIP is serviced cooperatively, request transmission and response
processing share the main application loop with DMA service. No blocking
socket call is used while a request is outstanding.

\begin{figure}[!t]
\centering
\includegraphics[width=\columnwidth]{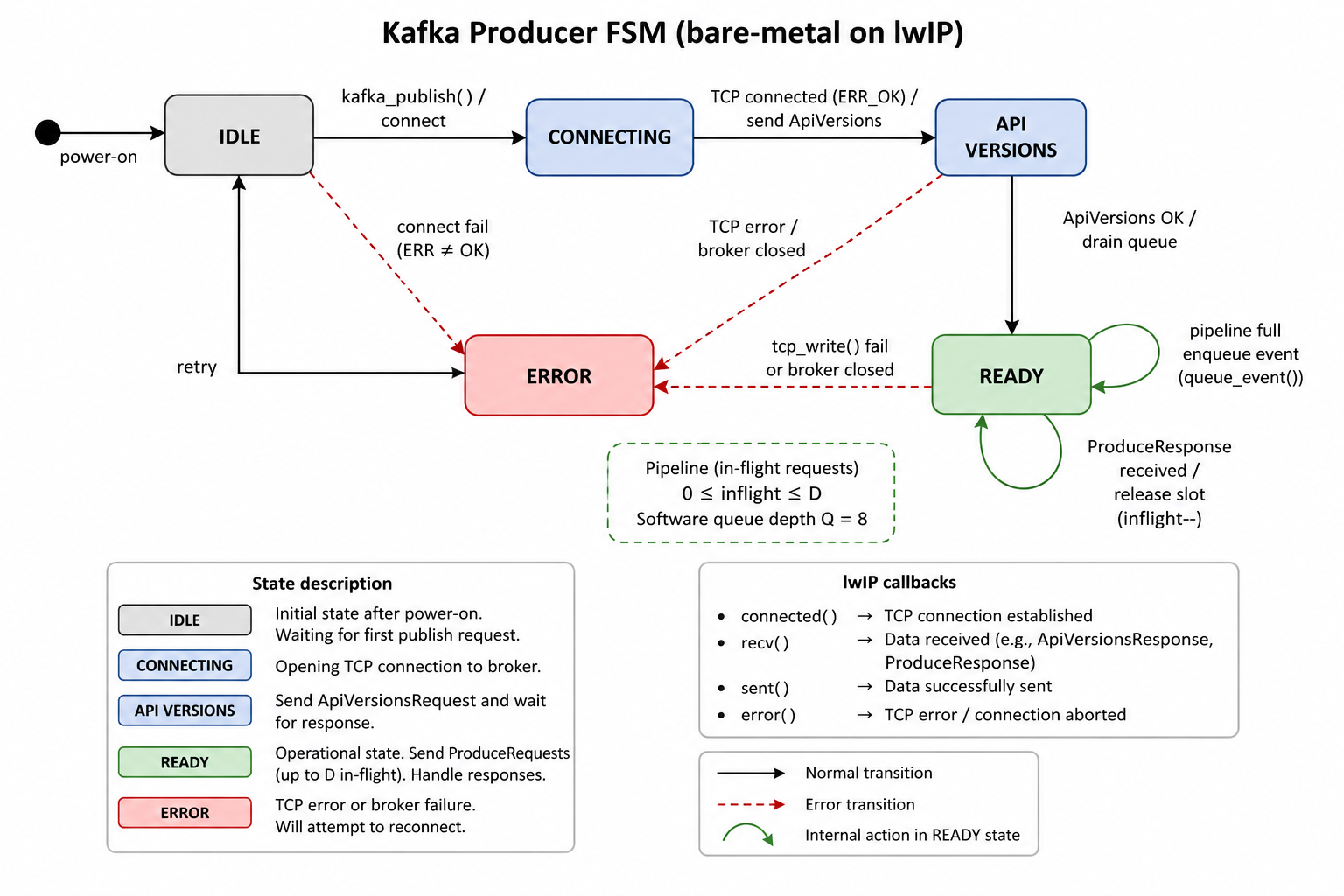}
\caption{Producer finite-state machine. In the ready state, up to $D$ Kafka
Produce requests may remain outstanding. A received Produce response releases
the associated request slot.}
\label{fig:fsm}
\end{figure}

\section{Pipelined Request Architecture}
\label{sec:pipeline}

\subsection{Pipeline Design}

The compile-time parameter $D$ sets the maximum number of Kafka
\texttt{ProduceRequest}s that may remain outstanding. Each in-flight request
uses a statically allocated slot and a unique correlation identifier.

For an idealized producer limited by a constant acknowledgement interval
$\tau_{\mathrm{ack}}$, the concurrency bound is

\begin{equation}
R_{\mathrm{ideal}} \approx \frac{D}{\tau_{\mathrm{ack}}}.
\label{eq:ideal_pipeline}
\end{equation}

Equation~\eqref{eq:ideal_pipeline} is not used as a prediction of the measured
rate. DMA service, request construction, TCP-buffer availability, software
queuing, network transport, broker processing, and response delivery are also
part of the measured path.

\subsection{Back-pressure and Software Queue}

The software queue depth was fixed at $Q=8$ in all measurements. If all $D$
in-flight slots are occupied, an event can be retained in this queue. A
queued event can advance after a \texttt{ProduceResponse} releases a slot.
The application continues to service lwIP while producer capacity is
unavailable.

Because $Q$ is constant and $D$ varies, the experiment changes request
concurrency and total producer capacity at the same time. The available state
is $D$ in-flight slots plus up to $Q$ queued events. The measured sweep is
therefore not a constant-total-buffer-capacity experiment.

\section{Experimental Methodology}
\label{sec:methodology}

\subsection{Network and Broker Configuration}

The Cora Z7-07S was connected through Gigabit Ethernet to a workstation
running an Apache Kafka broker in a Docker container. The Zynq board used
IP address 10.10.10.2 and the host used 10.10.10.1. The producer connected to
TCP port 9092. The event topic used one partition, replication factor one,
and \texttt{acks=1}.

Events were published to \texttt{zynq-events}. Each Kafka record contained the
8-byte payload defined in~\eqref{eq:event_format}. Measurement records were
kept in PS memory during acquisition. After all measured events in a run had
been acknowledged, the records were exported as CSV messages to
\texttt{zynq-metrics}. Metric-topic traffic was therefore absent from the
timed event-delivery phase.

Table~\ref{tab:experiment_config} collects the fixed settings used throughout
the measurement campaign.

\begin{table}[!t]
\renewcommand{\arraystretch}{1.08}
\caption{Experimental Configuration}
\label{tab:experiment_config}
\centering
\scriptsize
\begin{tabular}{ll}
\toprule
Parameter & Setting\\
\midrule
Board / device & Cora Z7-07S / XC7Z007S\\
Ethernet link & 1~Gb/s\\
Kafka Produce API & version~3\\
Acknowledgement policy & \texttt{acks=1}\\
Event topic & \texttt{zynq-events}, one partition\\
Application payload & 8~bytes\\
Software queue depth & $Q=8$\\
Pipeline depths & $D\in\{1,2,4,8\}$\\
Runs per depth & 20 (10 per source)\\
Events per run & 5000\\
\bottomrule
\end{tabular}
\end{table}

\subsection{Measurement Campaign}

The evaluated pipeline depths were

\begin{equation}
D\in\{1,2,4,8\}.
\end{equation}

Twenty independent runs were acquired at each depth. Source~1 and source~0
were alternated between successive runs, giving 10 runs per source and
pipeline depth. Each run contained 5000 events. The complete dataset contains

\begin{equation}
4\times20\times5000=400\,000
\end{equation}

acknowledged event metric records.

The run is the independent statistical unit. Event-level latency statistics
are first calculated within each 5000-event run and are then compared across
the 10 independent runs of each source and pipeline depth.

\subsection{Latency Definitions}

The PS-domain intervals are

\begin{align}
T_{\mathrm{DMA}} &=
 t_{\mathrm{DMA,done}}-t_{\mathrm{DMA,start}},
\label{eq:dma_latency}\\
T_{\mathrm{queue}} &=
 t_{\mathrm{send}}-t_{\mathrm{publish}},
\label{eq:queue_latency}\\
T_{\mathrm{send\rightarrow ack}} &=
 t_{\mathrm{ack}}-t_{\mathrm{send}},
\label{eq:send_ack_latency}\\
T_{\mathrm{publish\rightarrow ack}} &=
 t_{\mathrm{ack}}-t_{\mathrm{publish}}.
\label{eq:publish_ack_latency}
\end{align}

The PL timestamp is not combined with these PS timestamps. In addition,
$T_{\mathrm{send\rightarrow ack}}$ includes all time between local lwIP
acceptance and reception of the Kafka response; it is not a direct
measurement of wire latency or broker processing time.

\subsection{Throughput Definitions}

The steady acknowledgement rate is

\begin{equation}
R_{\mathrm{steady}}=
\frac{N-1}
{t_{\mathrm{ack},N-1}-t_{\mathrm{ack},0}},
\label{eq:steady_rate}
\end{equation}

and the operational rate is

\begin{equation}
R_{\mathrm{op}}=
\frac{N}
{t_{\mathrm{ack},N-1}-t_{\mathrm{DMA,start},0}},
\label{eq:operational_rate}
\end{equation}

with $N=5000$. The operational rate is used for the primary comparison. If a
run crosses the rollover of the 32-bit PS time counter, the counter is
unwrapped before the run-level interval is calculated.

\subsection{Statistical Treatment}

The run is treated as the independent experimental unit. For each source and
pipeline depth, 10 independent 5000-event runs are available. Event-level
latency statistics are first calculated separately within each run.

Run-level distributions were examined with the Shapiro--Wilk test. Several
pipeline groups departed from normality, and Levene's test indicated unequal
variance for operational throughput across pipeline depths (source~0:
$p=0.0126$; source~1: $p=0.0058$). Comparisons among pipeline depths were
therefore performed separately for each source with the Kruskal--Wallis test.
When the global test was significant, two-sided Mann--Whitney $U$ tests were
used for pairwise comparisons, with Holm correction applied to the six
pipeline-depth comparisons within each metric and source.

Kruskal--Wallis effect size is reported as $\epsilon^2$. Pairwise effect size
is reported as rank-biserial correlation. Mean operational throughput is
reported with a 95\% bootstrap confidence interval obtained by resampling the
independent runs. For latency, the median, P95, and P99 are calculated within
each run; the reported central value for each source and pipeline depth is the
median of the 10 run-level statistics. Bootstrap confidence intervals are also
computed across the independent runs.

\section{Experimental Results}
\label{Experimental}

\subsection{Run Integrity}

Each run was checked for record count, initial and final frame values,
frame-sequence continuity, and Kafka error codes before throughput or latency
statistics were calculated. Table~\ref{tab:integrity} gives the result.

\begin{table}[!t]
\renewcommand{\arraystretch}{1.15}
\caption{Integrity Checks for the Measurement Dataset}
\label{tab:integrity}
\centering
\scriptsize
\begin{tabular}{ccccc}
\toprule
$D$ & Runs & Records & Sequence errors & Kafka errors\\
\midrule
1 & 20 & 100\,000 & 0 & 0\\
2 & 20 & 100\,000 & 0 & 0\\
4 & 20 & 100\,000 & 0 & 0\\
8 & 20 & 100\,000 & 0 & 0\\
\midrule
Total & 80 & 400\,000 & 0 & 0\\
\bottomrule
\end{tabular}
\end{table}

Each run contained 5000 metric records, began at frame~0, ended at
frame~4999, and contained no sequence discontinuity in the exported metric
records. All recorded Kafka error codes were zero.

Two archived UART summaries, one for each source at $D=4$, also contain the
producer and network-stack counters printed at the end of the run. These
summaries are reported separately from the 80-run statistical dataset because
the same counters were not archived for every run. In both available summaries,
5000 DMA events produced 5000 publish calls, 5000 successful
\texttt{tcp\_write()} acceptances, and 5000 Kafka acknowledgements. Neither run
reported a dropped event, DMA error, Kafka error, lwIP \texttt{ERR\_MEM}, TCP
error, \texttt{tcp\_output()} error, unmatched response, or lost in-flight
request. Both summaries recorded 4996 queue admissions.

\begin{table}[!t]
\renewcommand{\arraystretch}{1.06}
\caption{Auxiliary Producer-Counter Validation at $D=4$}
\label{tab:counters_d4}
\centering
\scriptsize
\begin{tabular}{lrr}
\toprule
Counter & Source 0 & Source 1\\
\midrule
FPGA/DMA events & 5000 & 5000\\
Publish calls & 5000 & 5000\\
\texttt{tcp\_write()} accepted & 5000 & 5000\\
Queued & 4996 & 4996\\
Dropped & 0 & 0\\
Kafka ACK OK & 5000 & 5000\\
DMA / Kafka errors & 0 & 0\\
lwIP/TCP/output errors & 0 & 0\\
Unmatched / inflight lost & 0 & 0\\
\bottomrule
\end{tabular}
\end{table}

\subsection{Measured Throughput}

\begin{figure}[!t]
\centering
\includegraphics[width=\columnwidth]{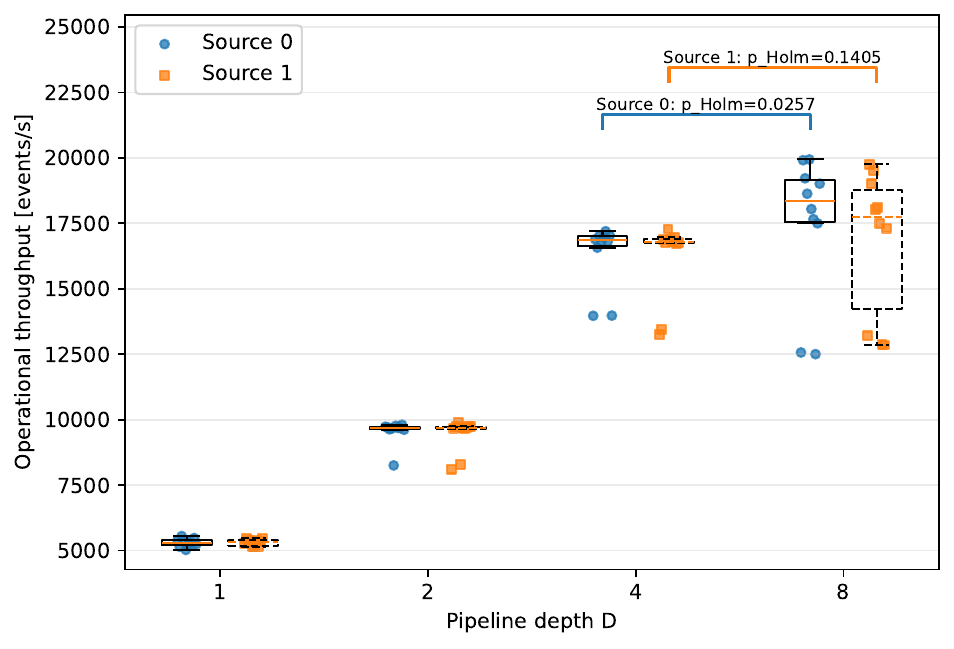}
\caption{Run-level operational throughput versus Kafka pipeline depth. Each
marker corresponds to one independent 5000-event run. Boxes summarize the 10
runs for each source. The bracket reports the Holm-adjusted pairwise
comparison between $D=4$ and $D=8$.}
\label{fig:throughput}
\end{figure}

Figure~\ref{fig:throughput} shows the operational throughput from
Eq.~\eqref{eq:operational_rate}.

At $D=1$, the mean rates were 5296 and 5314~events/s for sources~0 and~1.
At $D=2$, they were 9555 and 9424~events/s; at $D=4$, 16\,322 and
16\,179~events/s; and at $D=8$, 17\,499 and 16\,812~events/s.

Relative to $D=1$, the throughput gain

\begin{equation}
G(D)=\frac{R(D)}{R(1)}
\end{equation}

was 1.80 and 1.77 at $D=2$, 3.08 and 3.04 at $D=4$, and 3.30 and 3.16 at
$D=8$ for sources~0 and~1, respectively. From $D=4$ to $D=8$, the mean rate
increased by 7.2\% for source~0 and 3.9\% for source~1.

The concurrency efficiency

\begin{equation}
\eta(D)=\frac{R(D)}{D\,R(1)}
\end{equation}

was 0.90/0.89 at $D=2$, 0.77/0.76 at $D=4$, and 0.41/0.40 at $D=8$ for
sources~0/1. The measured rate therefore did not scale linearly with $D$ over
the evaluated range.

\subsection{Publish-to-Acknowledgement Latency}

Table~\ref{tab:measured_results} lists operational throughput and
publish-to-acknowledgement latency. Throughput entries are means with 95\%
bootstrap confidence intervals across the 10 independent runs. Latency entries
are the median of the 10 run-level median, P95, or P99 values.

\begin{table*}[!t]
\renewcommand{\arraystretch}{1.12}
\caption{Operational Throughput and Publish-to-Acknowledgement Latency}
\label{tab:measured_results}
\centering
\scriptsize
\begin{tabular}{cccccc}
\toprule
$D$ & Source &
$R_{\mathrm{op}}$ [events/s] &
Median [$\mu$s] &
P95 [$\mu$s] &
P99 [$\mu$s]\\
\midrule
1 & 0 & 5296 [5200, 5390] & 1640.0 & 1934.1 & 2186.0\\
1 & 1 & 5314 [5239, 5388] & 1616.0 & 1919.6 & 2157.0\\
2 & 0 & 9555 [9254, 9726] & 979.5  & 1185.5 & 1330.5\\
2 & 1 & 9424 [8969, 9743] & 981.0  & 1175.0 & 1319.0\\
4 & 0 & 16322 [15471, 16948] & 659.5 & 827.1 & 1078.5\\
4 & 1 & 16179 [15132, 16923] & 664.0 & 809.0 & 1046.6\\
8 & 0 & 17499 [15735, 18969] & 832.0 & 985.5 & 1297.1\\
8 & 1 & 16812 [15099, 18303] & 858.8 & 985.6 & 1267.7\\
\bottomrule
\end{tabular}
\end{table*}

The median publish-to-acknowledgement interval decreased from
1640/1616~$\mu$s at $D=1$ to 979.5/981~$\mu$s at $D=2$ and
659.5/664~$\mu$s at $D=4$ for sources~0/1. At $D=8$, it increased to
832/858.8~$\mu$s. The P95 values at $D=4$ were 827.1 and 809.0~$\mu$s; at
$D=8$ they were 985.5 and 985.6~$\mu$s.

\begin{figure}[!t]
\centering
\includegraphics[width=\columnwidth]{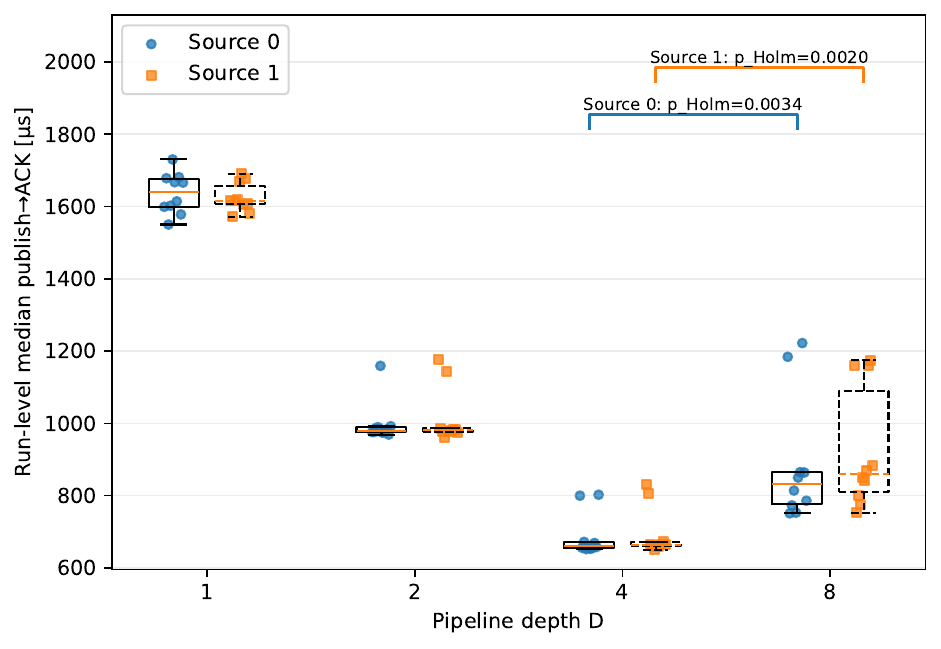}
\caption{Run-level median publish-to-acknowledgement interval versus pipeline
depth. Each marker corresponds to one independent run. Boxes summarize the 10
runs for each source. The bracket reports the Holm-adjusted $D=4$ versus
$D=8$ comparison.}
\label{fig:publish_ack_median}
\end{figure}

Among the evaluated depths, $D=4$ had the lowest median and P95
publish-to-acknowledgement values, while $D=8$ had the highest mean
operational throughput.

\subsection{Statistical Comparison Across Pipeline Depths}

Pipeline depth was associated with differences in operational throughput for
both sources (source~0: $H(3)=34.24$, $p=1.76\times10^{-7}$,
$\epsilon^2=0.868$; source~1: $H(3)=33.51$, $p=2.51\times10^{-7}$,
$\epsilon^2=0.848$). The run-level median publish-to-acknowledgement interval
also differed among pipeline depths (source~0: $H(3)=32.95$,
$p=3.31\times10^{-7}$, $\epsilon^2=0.832$; source~1: $H(3)=32.64$,
$p=3.83\times10^{-7}$, $\epsilon^2=0.823$). The corresponding P95 tests were
also significant for source~0 ($H(3)=31.94$, $p=5.39\times10^{-7}$) and
source~1 ($H(3)=31.07$, $p=8.22\times10^{-7}$).

For the pairwise comparison between $D=4$ and $D=8$, operational throughput
differed for source~0 ($p_{\mathrm{Holm}}=0.0257$, rank-biserial
$r=-0.60$) but not for source~1 ($p_{\mathrm{Holm}}=0.1405$, $r=-0.40$).
The median publish-to-acknowledgement interval increased between these depths
for both source~0 ($p_{\mathrm{Holm}}=0.0034$, $r=-0.84$) and source~1
($p_{\mathrm{Holm}}=0.0020$, $r=-0.88$). The P95 comparison gave
$p_{\mathrm{Holm}}=0.0182$ for source~0 and $0.0146$ for source~1. No
source~0 versus source~1 comparison was significant after Holm correction at
any pipeline depth for the reported throughput or publish-to-acknowledgement
metrics.

\subsection{Latency Decomposition}

\begin{figure}[!t]
\centering
\includegraphics[width=\columnwidth]{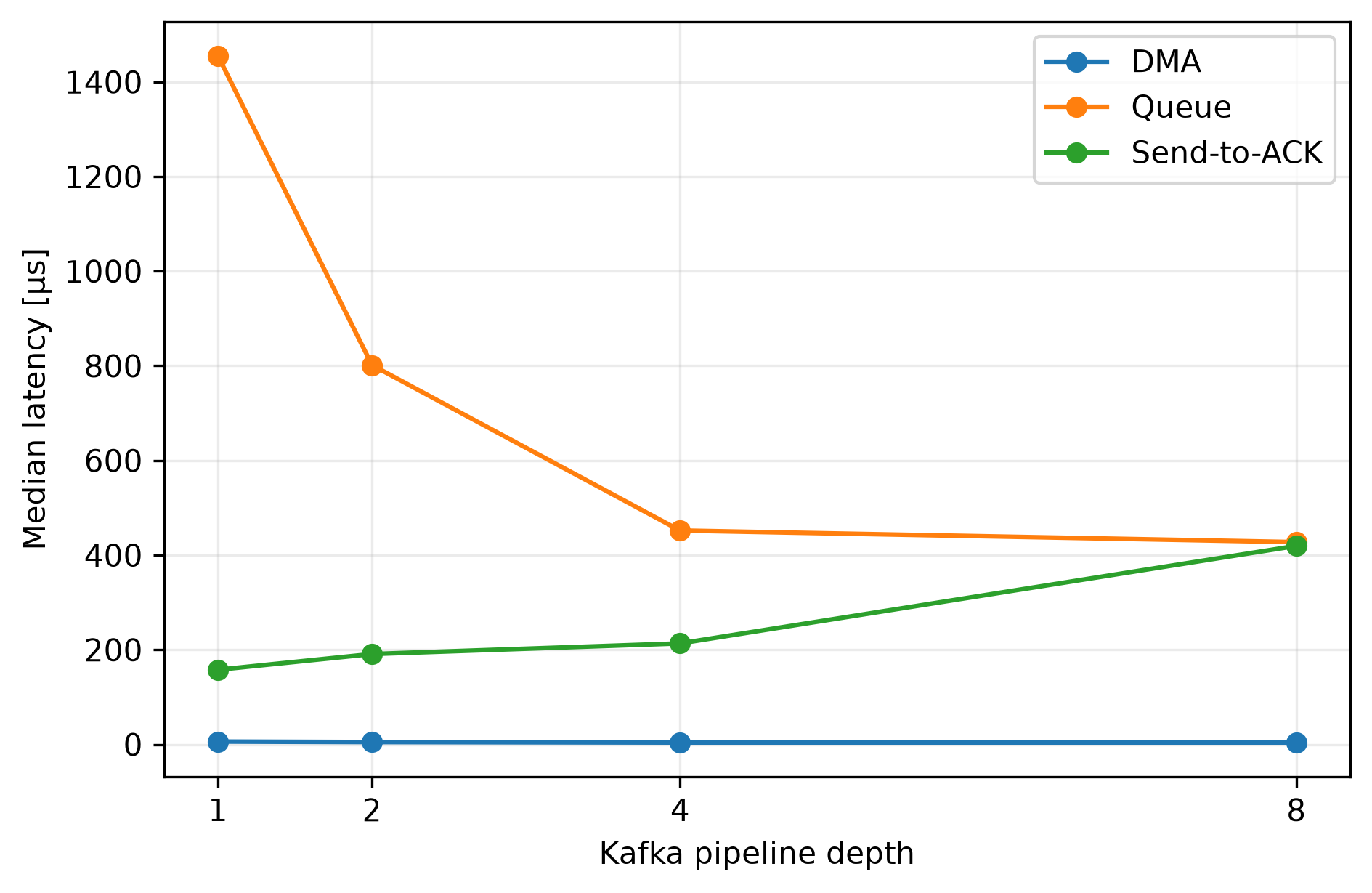}
\caption{Run-level median latency components versus pipeline depth. Components
are plotted separately because their medians are not additive.}
\label{fig:latency_components}
\end{figure}

Figure~\ref{fig:latency_components} plots the DMA, software-queue, and
send-to-acknowledgement intervals separately. The curves are not stacked
because medians of the component distributions are not additive.

The median DMA interval was 6--7~$\mu$s at $D=1$, 5~$\mu$s at $D=2$,
4~$\mu$s at $D=4$, and 4--5~$\mu$s at $D=8$.

For source~0, the median queue interval was 1476~$\mu$s at $D=1$,
801~$\mu$s at $D=2$, 450~$\mu$s at $D=4$, and 418.5~$\mu$s at $D=8$.
For source~1, the corresponding values were 1455, 801.5, 452, and
432.5~$\mu$s.

The median send-to-acknowledgement interval was approximately
158~$\mu$s at $D=1$, 191--192~$\mu$s at $D=2$, 213.5--214~$\mu$s at
$D=4$, and 410--424.5~$\mu$s at $D=8$. Thus, from $D=4$ to $D=8$, the
queue interval decreased while the send-to-acknowledgement interval
increased.

The instrumentation does not timestamp actual Ethernet transmission and does
not separate network transport, broker processing, and response delivery.
The increase in $T_{\mathrm{send\rightarrow ack}}$ at $D=8$ is therefore not
assigned to a specific mechanism.

\subsection{Measurement Scope}

The measurements use finite runs of 5000 events. They do not characterize
long-duration saturation behavior. The PL timestamp is included in every
event but is not combined with the PS timestamps because a common clock origin
has not been established.

The integrity checks in Table~\ref{tab:integrity} apply to the acknowledged
event metric records exported by the producer. They do not constitute an
independent host-side decode of the binary \texttt{zynq-events} payload.

\section{Post-Implementation Results}
\label{sec:implementation_results}

The implementation reports were generated with Vivado~2024.2 for the
XC7Z007S device. The PL clock \texttt{clk\_fpga\_0} has a constrained period
of 7.692~ns, corresponding to 130.005~MHz. Post-implementation timing analysis
reported a worst negative slack (WNS) of 0.982~ns and total negative slack
(TNS) of 0~ns over 13\,087 setup endpoints. The worst pulse-width slack was
2.596~ns with no failing pulse-width endpoints. Vivado reported that all
user-specified timing constraints were met.

\begin{table}[!t]
\renewcommand{\arraystretch}{1.08}
\caption{Post-Implementation Timing Summary}
\label{tab:timing}
\centering
\scriptsize
\begin{tabular}{lr}
\toprule
Quantity & Value\\
\midrule
PL clock period & 7.692~ns\\
PL clock frequency & 130.005~MHz\\
WNS & 0.982~ns\\
TNS & 0~ns\\
Failing setup endpoints & 0 / 13\,087\\
Worst pulse-width slack & 2.596~ns\\
Failing pulse-width endpoints & 0 / 5077\\
\bottomrule
\end{tabular}
\end{table}

The hierarchical utilization report attributes most PL logic to the two AXI
DMA blocks and the AXI interconnect infrastructure. Table~\ref{tab:hier_resources}
lists the main acquisition-related blocks. The complete design uses 3203 LUTs,
4758 flip-flops, ten RAMB36 blocks, and four RAMB18 blocks. The two event-source
blocks use six LUTs and 97 flip-flops each.

\begin{table}[!t]
\renewcommand{\arraystretch}{1.06}
\caption{Hierarchical PL Resource Utilization}
\label{tab:hier_resources}
\centering
\scriptsize
\begin{tabular}{lrrrr}
\toprule
Block & LUT & FF & RAMB36 & RAMB18\\
\midrule
AXI DMA 0 & 1007 & 1570 & 4 & 1\\
AXI DMA 1 & 1007 & 1570 & 4 & 1\\
AXI interconnect & 171 & 215 & 0 & 0\\
AXIS FIFO 0 & 48 & 53 & 1 & 1\\
AXIS FIFO 1 & 48 & 53 & 1 & 1\\
AXI GPIO 0 & 40 & 49 & 0 & 0\\
AXI GPIO 1 & 37 & 39 & 0 & 0\\
Event source 0 & 6 & 97 & 0 & 0\\
Event source 1 & 6 & 97 & 0 & 0\\
\midrule
Complete design & 3203 & 4758 & 10 & 4\\
\bottomrule
\end{tabular}
\end{table}

The device-level utilization summary corresponds to 22\% of LUTs, 3\% of
LUTRAM, 17\% of flip-flops, 24\% of BRAM, 2\% of I/O resources, and 3\% of
BUFG resources. Vivado estimated total on-chip power at 1.54~W, comprising
1.422~W dynamic and 0.119~W static power. The power report labels its
confidence level as \emph{Low}; these values are tool estimates and were not
obtained from an external power measurement.

\section*{Code Availability}

The source code for the bare-metal Kafka producer and the analysis scripts
used in this work are available at
\url{https://github.com/feZapBas/apache_kafka_baremetal_producer}.
\balance
\section{Conclusions}

A bare-metal Kafka producer was implemented on a Zynq-7000 SoC using the
Kafka Wire Protocol over the lwIP raw TCP API. FPGA-generated events are
transferred through AXI DMA and published without an intermediate gateway
computer. Each 8-byte event contains a 32-bit frame counter and a 32-bit PL
timestamp. The PS records DMA, publication, local TCP-acceptance, and Kafka
response times.

The measurement campaign contains 80 independent 5000-event runs for
$D\in\{1,2,4,8\}$ and two DMA sources. All 400\,000 acknowledged metric
records passed the run-level sequence and Kafka-error checks. Mean operational
throughput increased from approximately 5.3~kevents/s at $D=1$ to
approximately 16.2~kevents/s at $D=4$ and 16.8--17.5~kevents/s at $D=8$.

The lowest measured median publish-to-acknowledgement interval occurred at
$D=4$: 659.5~$\mu$s for source~0 and 664~$\mu$s for source~1. The P95 values
were 827.1 and 809.0~$\mu$s. At $D=8$, the median increased to 832 and
858.8~$\mu$s while the mean operational rate increased to 17\,499 and
16\,812~events/s. The $D=4$ versus $D=8$ throughput difference was detected
for source~0 ($p_{\mathrm{Holm}}=0.0257$) but not for source~1
($p_{\mathrm{Holm}}=0.1405$), whereas the increase in median
publish-to-acknowledgement interval was detected for both sources
($p_{\mathrm{Holm}}=0.0034$ and $0.0020$). From $D=4$ to $D=8$, the measured
queue interval decreased and the send-to-acknowledgement interval increased.

The implemented PL design operates with a 7.692~ns clock constraint
(130.005~MHz) and met all reported post-implementation timing constraints,
with WNS of 0.982~ns and TNS of 0~ns. Hierarchical utilization shows that the
two AXI DMA blocks account for 2014 of the 3203 LUTs used by the complete
design.

The PL and PS time domains remain separate in this analysis. The current
instrumentation also does not separate the interval after local TCP acceptance
into Ethernet-transmission, network, broker-processing, and response-delivery
components. The run-integrity analysis is based on producer-exported metric
records rather than an independent host-side decode of the binary event
payload.

\balance



\begin{thebibliography}{10}

\bibitem{kafka_original}
J.~Kreps, N.~Narkhede, and J.~Rao,
``Kafka: A distributed messaging system for log processing,''
in \emph{Proc. NetDB Workshop}, 2011.

\bibitem{kafka_protocol}
Apache Software Foundation,
``Kafka Protocol Guide,''
\url{https://kafka.apache.org/protocol.html}.

\bibitem{zynq_trm}
AMD-Xilinx,
``Zynq-7000 SoC Technical Reference Manual,''
UG585.

\bibitem{cora_refman}
Digilent Inc.,
``Cora Z7 Reference Manual,''
\url{https://digilent.com/reference/programmable-logic/cora-z7/reference-manual}.

\bibitem{lwip_dunkels}
A.~Dunkels,
``Design and implementation of the lwIP TCP/IP stack,''
Swedish Institute of Computer Science, 2001.

\bibitem{xilinx_lwip}
AMD-Xilinx,
``Lightweight IP Application Examples,''
XAPP1026.

\bibitem{crc32c_castagnoli}
G.~Castagnoli, S.~Br\"auer, and M.~Herrmann,
``Optimization of cyclic redundancy-check codes with 24 and 32 parity bits,''
\emph{IEEE Trans. Commun.}, vol.~41, no.~6, pp.~883--892, Jun.~1993.

\bibitem{mqtt_embedded}
A.~Stanford-Clark and H.~L.~Truong,
``MQTT for sensor networks (MQTT-SN) version 1.2,''
OASIS Standard, 2013.

\bibitem{amqp_embedded}
OASIS,
``Advanced Message Queuing Protocol (AMQP) v1.0,''
OASIS Standard, 2012.

\bibitem{auger_amiga}
A.~Aab \emph{et al.} (Pierre Auger Collaboration),
``The Pierre Auger Observatory: Contributions to the 36th ICRC,'' 2019.

\bibitem{andes_detector}
X.~Bertou \emph{et al.},
``The ANDES deep underground laboratory,''
\emph{Eur. Phys. J. Plus}, vol.~135, 2020.

\end{thebibliography}
\end{document}